\documentclass[aps,prl,showpacs,twocolumn,groupedaddress]{revtex4}

\usepackage{amssymb, amsmath}
\usepackage{graphicx}

\begin{document}


\title{Determinable Solutions for One-dimensional Quantum Potentials:
Scattering, Quasi-bound and Bound State Problems}


\author{Hwasung Lee}
\email[]{hl329@cam.ac.uk}
\affiliation{Department of Physics and Astronomy, Seoul National
University, Seoul 151-747, Korea}
\altaffiliation{Current address: Centre for Mathematical Science, Wilberforce Road, University of Cambridge, Cambridge CB3 0WA, United Kingdom}

\author{Y. J. Lee}
\email[]{yjlee@dankook.ac.kr} \affiliation{Department of Physics,
Dankook University, Cheonan 330-714, Korea}


\date{\today}

\begin{abstract}
We derive analytic expressions of the recursive solutions to the
Schr\"{o}dinger's equation by means of a cutoff potential technique
for one-dimensional piecewise constant potentials. These solutions
provide a method for accurately determining the transmission
probabilities as well as the wave function in both classically
accessible region and inaccessible region for any barrier
potentials. It is also shown that the energy eigenvalues and the
wave functions of bound states can be obtained for potential-well
structures by exploiting this method. Calculational results of
illustrative examples are shown in order to verify this method for
treating barrier and potential-well problems.
\end{abstract}


\maketitle


\section{I. Introduction}
Quantum mechanical tunneling and quantum well problems have
attracted more interest, due to recent advances in the fabrication
of semiconductor layers and the development of high-speed and novel
devices \cite{Barnham}. In order to better understand the physical
properties of a device, it is important that one can accurately
solve and analyze the one-dimensional potential problems. A number
of methods for solving the Schr\"{o}dinger's equation and for
locating the bound states and resonances generated by
one-dimensional potentials have been developed over the past decades
\cite{Brennan,Ghatak,Jonsson,Glytsis,Rakityansky}. Most of them are
based on the so-called transfer-matrix approach. There are also some
methods that are cumbersome to implement, for instance, the Monte
Carlo method \cite{Singh} and the finite element method (FEM)
\cite{Nakamura}.

In contrast to those numerical methods, a limited number of exact
analytic solutions are available only for simple potential
structures; analytic solutions are preferred due to its simple form
and clear interpretation of the physics underlying the process. WKB
method, as an approximate analytic approach, has been widely used
but it is restricted to slowly varying potential profiles that are
continuous. Improved methods such as the modified conventional WKB
(MWKB)\cite{Love} and the modified Airy functions
(MAF)\cite{Ghatak0} still fail to provide perfect results.

Tikochinsky\cite{Tikochinsky} has derived two nonlinear first-order
equations replacing the second-order linear Schr\"{o}dinger's
equation in a cutoff potential method for one-dimensional scattering
amplitudes. In this paper we solve these nonlinear first-order
equations analytically and obtain recursive solutions of the
Schr\"{o}dinger's equation for multi-step barrier potentials. We
intend to find a method providing a general analysis of the
one-dimensional problem for both classically accessible region and
inaccessible region. In fact, arbitrarily accurate solutions
including the scattering amplitudes and the wave functions are
obtainable with these recursive solutions for any potential profile
by dividing it into many segments, since any continuous potential
problem can be recovered as the segments become finer and finer.

Using the recursive solutions, we will also determine the wave
functions of quasi-bound and bound states. We can solve a
potential-well problem by evenly uplifting the potential function
restricted on the potential-well region and its surroundings to
construct a resonant barrier-potential that can readily be handled
as a model potential. The resonant barrier-potential constitutes a
quasi-bound-state problem, therefore each sharp tunneling resonance
is locatable so that a quasi-bound-state eigenvalue can be
determined which immediately leads to a bound-state eigenvalue being
sought for the original potential well. Then, we can determine the
eigenfunction belonging to the quasi-bound-state eigenvalue by means
of the recursive solutions.

In Section II, we review the basic formalism for the one-dimensional
problems that allows two nonlinear first-order equations to replace
the linear second-order Schr\"{o}dinger's equation. In section III,
we derive analytic expressions with recursive coefficients by
solving the nonlinear equations for the transmission amplitudes,
reflection amplitudes, and the wave functions for piecewise constant
potentials. In section IV, a quasi-bound potential profile which
characterizes a resonant tunneling is considered as the model
potential of a potential well in order to treat bound-state
problems. In section V and VI, we demonstrate the validity of the
method by taking some examples, and discuss the calculational
results.


\section{II. Basic Formalism}
The one-dimensional Schr\"{o}dinger equation for a particle with
mass $m$ incident upon a potential energy $U(z)$ that asymptotically
vanishes can be written as
\begin{equation}
\label{Schr} \left( {{d^2}\over{dx^2}} + k^2 \right) \psi(x) = V(x)
\psi(x)
\end{equation}
where $x=z\sqrt{2m\epsilon /\hbar^2}$, $k^2=K/\epsilon$, and
$V(x)=U(z)/\epsilon$ with a certain unit energy $\epsilon$ chosen
for convenience. Here, $z$ is the position of the particle in the
one-dimensional coordinate, $K$ is the energy of the particle, and
$\hbar=h/2\pi$, $h$ being Planck's constant. Thus, the coordinate $x$,
the wave number $k$ and the potential $V(x)$ are all dimensionless.
The dimensionless energy $E$ is defined by
\begin{eqnarray}
E=k^2.
\end{eqnarray}
Now. we introduce a cutoff potential that was used in
Refs.~\onlinecite{Tikochinsky,Glen1, Glen2, Razavy}. The cutoff
potential is defined by
\begin{eqnarray}
\label{cutoffpot} V_c(y,x) = V(x)\theta(x-y),
\end{eqnarray}
where $\theta(x-y)$ is the step function. If $V(x)$ is replaced by
the cutoff potential, Eq.~\eqref{Schr} has the formal solution:
\begin{equation}
\label{cutoffwave} \psi_E(y,x)=A e^{ikx} + {{1}\over{2ik}} \int
e^{ik|x-x'|} V_c(y,x') \psi_E (y,x')dx'.
\end{equation}
Then, the cutoff reflection and cutoff transmission amplitudes for
this problem can, respectively, be written as
\begin{eqnarray}
\label{cutoffR} R_E(y)= {{1}\over{2ikA}} \int e^{ikx'}
V_c(y,x')\psi_E(y,x')dx'
\end{eqnarray}
and
\begin{eqnarray}
\label{cutoffT} T_E(y)=1+ {{1}\over{2ikA}} \int  e^{-ikx'}
V_c(y,x')\psi_E(y,x')dx'.
\end{eqnarray}

In Ref.~\onlinecite{Tikochinsky} it has been shown that the problem
can be stated as two nonlinear first order equations given by
\begin{eqnarray}
\label{derivR} {{dR_E(y)}\over{dy}}= -{{1}\over{2ik}} V(y)
\left[e^{iky}+e^{-iky} R_E(y)\right]^2
\end{eqnarray}
and
\begin{eqnarray}
\label{derivT} {{dT_E(y)}\over{dy}} = -{{e^{-iky}}\over{2ik}} V(y)
\left[e^{iky}+e^{-iky} R_E(y)\right]T_E(y),
\end{eqnarray}
with the boundary conditions, $R_E(\infty)=0$ and $T_E(\infty)=1$.
The Schr\"{o}dinger's equation, which is a second-order differential
equation, has been disassembled through the cutoff-potential
manipulation into two first-order equations, Eqs.~\eqref{derivR} and
\eqref{derivT}. Integration of Eq. \eqref{derivT} immediately leads to
\begin{equation}
\label{Tfunc} T_E(x)=\exp\left(\int_x^\infty {{e^{-iky}}\over{2ik}}
V(y)\left[ e^{iky}+e^{-iky}R_E(y)\right]dy \right).
\end{equation}
The reflection amplitude $R_E$ and the transmission amplitude $T_E$
for the potential $V(x)$ are $R_E=R_E(-\infty)$ and
$T_E=T_E(-\infty)$ in terms of which the reflection and transmission
probabilities are, respectively, given as
\begin{eqnarray}
P_R=\left| R_E\right|^2\\
P_T={{k_f}\over{k}}\left| T_E\right|^2,
\end{eqnarray}
where $k_f=\sqrt{E-V(\infty)}$.

\section{III. The Piecewise-constant Potential Problem}
We want to find analytic expressions of the transmission amplitude,
reflection amplitude and the wave function for a multi-step
potential profile which contains $n$ layers of constant potential in
the zero-potential environment, as shown in Fig.~\protect\ref{one}.
Let us solve Eq.~(\ref{derivR}) for $R_E(x)$ in $j$th step region
($b_{j-1}<x<b_j$) for $V(x)=V_j\neq 0$, where $V_j$ is constant for
$j=1,\cdots,n$. By defining
\begin{eqnarray} Q_E(x)= R_E(x) e^{-2ikx} + 1 -
{{2k^2}\over{V_j}}
\end{eqnarray}
Eq.~(\ref{derivR}) can be rewritten as
\begin{eqnarray}
\label{Q} {{dQ_E(x)}\over{dx}} = i {{V_j}\over{2k}} \left[
\left(Q_E(x)\right)^2+\left({{4k^2}\over{V_j^2}}(V_j-k^2)\right)\right]
\end{eqnarray}
to which the solution is
\begin{eqnarray}
Q_E(x)= i2k {{p_j}\over{V_j}}
\left({{1+A_je^{-2p_jx}}\over{1-A_je^{-2p_jx}}}\right)
\end{eqnarray}
with a constant $A_j$ to be determined by a boundary condition.
Here, $p_j$ is defined by
\begin{eqnarray}
p_j=\left\{ \begin{array}{ll} \sqrt{V_j-k^2}  &\mbox{for $V_j>k^2$},\\
  i\sqrt{k^2-V_j} & \mbox{for $V_j<k^2$}.
  \end{array}
  \right.
\end{eqnarray}
Then, we
obtain the cutoff reflection amplitude in the $j$th step region:
\begin{equation} \label{Rpiecesol} R_E(x)=\left[
{{2k^2}\over{V_j}}-1 + i 2k{{p_j}\over{V_j}}\left( {{1+A_je^{-2p_j
x}}\over{1-A_je^{-2p_j x}}}\right) \right] e^{2ikx}.
\end{equation}
On the other hand, the solution of Eq. \eqref{derivR} for $V_j=0$ in
the $j$th step region is
\begin{eqnarray}
R_E(x) = C_j
\end{eqnarray}
which can be determined by the continuity condition of $R_E(x)$ at
the boundary $x=b_j$.

\begin{figure}
\centering
\begin{picture}(170,120)(0,0)
\put(10,10){\vector(0,1){100}} \put(10,10){\vector(1,0){160}}

\put(20,10){\line(0,1){30}} \put(20,40){\line(1,0){20}}
\put(20,0){$b_0$} \put(25,45){$V_1$}

\put(40,10){\line(0,1){60}} \put(40,70){\line(1,0){20}}
\put(40,0){$b_1$} \put(45,75){$V_2$}

\put(60,10){\line(0,1){60}} \put(70,40){$\cdots$}
\put(60,0){$b_2$}

\put(90,10){\line(0,1){80}} \put(90,90){\line(1,0){20}}
\put(90,0){$b_{j-1}$} \put(95,95){$V_j$}

\put(110,10){\line(0,1){80}} \put(120,40){$\cdots$}
\put(110,0){$b_j$}

\put(140,10){\line(0,1){40}} \put(140,50){\line(1,0){20}}
\put(140,0){$b_{n-1}$}

\put(160,10){\line(0,1){40}} \put(160,0){$b_n$} \put(145,55){$V_n$}
\end{picture}
\caption{\label{one}} A multi-step potential profile consisting of $n$ steps in the zero-potential environment.
\end{figure}

If $V_{j+1}=0$, $R_E(x)=C_{j+1}$ in the region $b_j<x<b_{j+1}$
and the continuity of $R_E(x)$ at the boundary $x=b_j$ leads to
\begin{equation}
\label{AjCjp1}
A_j=\left({{V_j(C_{j+1}e^{-2ikb_j}+1)-2k^2-i2kp_j}
\over{V_j(C_{j+1}e^{-2ikb_j}+1)-2k^2+i2kp_j}}
\right)e^{2p_jb_j}.
\end{equation}
But, if $V_{j+1}\neq 0$, the boundary condition gives
\begin{eqnarray}
\label{AjAjp1} A_j=\left( {{D_j-V_{j+1}p_j+i
k(V_{j+1}-V_j)}\over{D_j+V_{j+1}p_j+i k(V_{j+1}-V_j)}} \right)
e^{2p_jb_j},
\end{eqnarray}
where
\begin{eqnarray}
\label{Dj} D_j =V_j\, p_{j+1} \left( {{1+A_{j+1}
e^{-2p_{j+1}b_j}}\over{1-A_{j+1} e^{-2p_{j+1}b_j}}} \right).
\end{eqnarray}
Let us take the index $j=n+1$ for the region $x>b_n$. Since
$V_{n+1}=0$ and $p_{n+1}=ik$, we obtain $A_{n+1}=0$ from
\eqref{AjCjp1}. $A_j$ is related to $A_{j+1}$ through
Eq.~\eqref{AjAjp1} if $V_{j+1}\neq 0$, while it is related to the
constant reflection amplitude $C_{j+1}$ through Eq.~\eqref{AjCjp1}
if $V_{j+1}=0$. Therefore, all $A_j$'s for $j=1,\cdots,n$ can be
determined recursively from the starting value $A_{n+1}=0$.

The cutoff transmission amplitude $T_E(x)$ can be obtained
in the $j$th step region ($b_{j-1}<x<b_j$) from Eq.~(\ref{Tfunc})
by the help of
\eqref{Rpiecesol}.
\begin{eqnarray}
\label{Tpiecesol} T_E(x)&=&T_E(b_j)\exp\left[ (p_j-ik)(b_j-x)\right]\nonumber\\
&&\times\left( {{1-A_j e^{-2p_jb_j}}\over{1-A_j e^{-2p_jx}}}
\right)\end{eqnarray} where
\begin{eqnarray}
T_E(b_j)&=&\prod_{l=j+1}^n\exp\left[ (p_l-i k)(b_l-b_{l-1})\right]\nonumber\\
&&\times\left( {{1-A_l e^{-2p_lb_l}}\over{1-A_l e^{-2p_lb_{l-1}}}}
\right).
\end{eqnarray}
 Therefore, the analytic expression of the transmission amplitude is
\begin{eqnarray}
\label{Tpiecesolt} T_E&=&T_E(-\infty)=\prod_{l=1}^n\exp\left[ (p_l-i
k)(b_l-b_{l-1})\right] \nonumber\\
&&\times\left( {{1-A_l e^{-2p_lb_l}}\over{1-A_l
e^{-2p_lb_{l-1}}}} \right).
\end{eqnarray}

From Eqs.~\eqref{cutoffwave} and \eqref{cutoffR}
\begin{eqnarray}
\psi_E(y,y)=A\left[e^{iky}+e^{-iky}R_E(y)\right]
\end{eqnarray}
and by differentiating Eq.~\eqref{cutoffwave} in the region $y<x$
one has the relation
\begin{eqnarray}
\label{derivPsi1} {{\partial\psi_E(y,x)}\over{\partial y}}&=&
-{{e^{-iky}V(y)}\over{2ik}} \psi_E(y,y)e^{ikx}\\
&+& {{1}\over{2ik}}\int
e^{ik|x-x'|}V_c(y,x') {{\partial \psi_E(y,x')}\over{\partial y}}dx'.\nonumber
\end{eqnarray}
Eq.~\eqref{cutoffwave} and Eq.~\eqref{derivPsi1} are both the
solution of Eq.~\eqref{Schr} with the cutoff potential $V_c(y,x)$
but have different amplitudes. Their amplitudes differ by a factor
$-e^{-iky}V(y) \psi_E(y,y)/2ikA$, thus in the region $y<x$
\begin{equation}
\label{derivPsi2} {{\partial\psi_E(y,x)}\over{\partial y}} =
-{{e^{-iky}V(y)}\over{2ik}}\left[e^{iky}+e^{-iky}R_E(y)\right]\psi_E(y,x).
\end{equation}
Integration of Eq.~\eqref{derivPsi2} in the region $y<x$ leads to
\begin{eqnarray}
&&{{\psi_E(x,x)}\over{\psi_E(x)}}={{\psi_E(x,x)}\over{\psi_E(-\infty,x)}}\\
&=&
\exp \left(-\int_{-\infty}^x {{e^{-iky}}\over{2ik}}V(y)
\left[e^{iky}+e^{-iky}R_E(y)\right]dy \right)\nonumber
\end{eqnarray}
which can be rewritten, with the aid of Eq.~\eqref{Tfunc}, as
\begin{eqnarray}
\label{solwave} \psi_E(x)=A
{{T_E}\over{T_E(x)}}\left[e^{ikx}+e^{-ikx} R_E(x)\right].
\end{eqnarray}
One can imagine the insertion of an infinitesimally thin virtual
layer of zero potential at $x$, and consider the wave function
$\psi_E(x)$ to be the superposition of the forward traveling wave
$e^{ikx}$ and the backward traveling wave $R_E(x)~e^{-ikx}$
reflected upon the potential on the right side of the virtual layer,
multiplied by the factor $AT_E/T_E(x)$ which is fathomable to be the
correct amplitude for the superposed wave function at $x$.

Substituting \eqref{Tpiecesol} and \eqref{Tpiecesolt} into
Eq.~\eqref{solwave}, we write the wave function for the energy $E$
in the $j$th layer ($b_{j-1}<x<b_j$) as
\begin{widetext}
\begin{eqnarray}
\psi_{jE}(x)&=&A \prod_{l=1}^{j}\exp\left[ (p_l-i
k)(b_l-b_{l-1})\right] \left( {{1-A_l e^{-2p_lb_l}}\over{1-A_l
e^{-2p_lb_{l-1}}}}\right) \nonumber\\ &\times&\exp\left[ (p_j-i
k)(x-b_j)\right] \left( {{1-A_j e^{-2p_jx}}\over{1-A_j
e^{-2p_jb_j}}}\right)\left[e^{ikx}+e^{-ikx}R_E(x)\right]\nonumber\\ 
&=&A\prod_{l=1}^{j}\exp\left[ (p_l-i k)(b_l-b_{l-1})\right] 
\left({{1-A_l e^{-2p_lb_l}}\over{1-A_l e^{-2p_lb_{l-1}}}}\right)\nonumber\\ &\times&
{{2ik\exp\left[(ik-p_j)b_j\right]}
\over{V_j\left[1-A_j\exp\left(-2p_jb_j\right)\right]}}
\left[\left(-ik+p_j\right) e^{p_jx} +A_j\left(ik+p_j\right)
e^{-p_jx}\right]. \label{analypsi}
\end{eqnarray}
\end{widetext}
Eq.~\eqref{analypsi} is an analytic expression for the recursive
solution of the Schr\"{o}dinger equation. It is simple and fast to
calculate $A_j$'s for any multi-step potential. Also, any continuous
potential problem may be considered to be a multi-step potential
problem by dividing the non-zero potential region into a lot of
small segments such that the potential in each segment can be
approximated to a specific value, for instance, the average value,
of the continuous potential within the segment. Thus, the analytic
expression of the wave function given by Eq.~\eqref{analypsi} can be
used for obtaining an arbitrarily accurate solution for any
one-dimensional potential.

In passing, let us check that Eq.~\eqref{analypsi} is consistent
with the traveling plane waves which are expected in the regions
$x<b_0$ and $x>b_n$, where $V(x)=0$. Without loss of generality, one
may take the first step region ($b_0<x<b_1$) such that $V_1=0$.
Then, the wave function in the region $x<b_1$ becomes
\begin{align}\label{inwave}
\psi_{1E}(x)=& \lim_{V_1\rightarrow 0}A(e^{ikx}+e^{-ikx}R_E(x))\nonumber\\
& \times\exp[(p_1-ik)(x-b_0)]\frac{1-A_1 e^{-2p_1 x}}{1-A_1 e^{-2p_1b_0}}\nonumber\\
=& A\left[e^{ikx}+R_E ~e^{-ikx}\right],
\end{align}
since $A_1=-(V_1/4k^2)R_E$ from Eq.~(\ref{Rpiecesol})
for the limit $V_1 \rightarrow 0$. On the other hand, if we take the
last step region $b_{n-1}<x<b_n$ such that $V(x)=V_n=0$, the
wavefunction in the region  $x>b_{n-1}$ becomes
\begin{eqnarray}
\label{outwave} \psi_{nE}(x)=A T_E ~e^{ikx},
\end{eqnarray}
since $R_E(x)=0$ and $A_n=-(V_n/4k^2)R_E(x)$ from
Eqs.~(\ref{Rpiecesol}) and (\ref{Tpiecesolt}) in the limit $V_n
\rightarrow 0$. Eqs.~\eqref{inwave} and \eqref{outwave} show that
Eq.~\eqref{analypsi} is consistent as the solution of
Eq.~\eqref{Schr} in the zero-potential region.

\section{IV. Quasi-bound States}
We consider a resonant barrier structure that consists of $n$ layers
of piecewise constant potential. Resonances are supposed to occur at
energies less than both $V_1$ and $V_n$ which are potential values
of the end steps so that quasi-bound states are involved. The wave
function $\psi_{jE}(x)$ in $j$th layer for this structure can be
obtained with the calculated $A_j$'s which are dependent on $E$.
$A_n$, in the last layer, can be evaluated from \eqref{AjCjp1} with
cutoff reflection amplitude $C_{n+1}=0$. Inserting the evaluated
$A_n$ into \eqref{analypsi}, we find in the last layer for $E<V_n$
\begin{eqnarray}
\label{lastBarrier} \psi_{nE}(x)&\propto& (-ik+p_n)e^{p_n x} \\
&+&
{{V_n-2k^2-2ikp_n}\over{V_n-2k^2+2ikp_n}} (ik+p_n) e^{p_n(2b_n-x)},\nonumber
\end{eqnarray}
where the first term exponentially increases with increasing $x$,
while the second term exponentially decreases. One can see that if
$p_n(b_n-b_{n-1})$ is large enough, the first term is negligible in
magnitude compared with the second term, so the wave function
exponentially damps with increasing $x$ in the region
$b_{n-1}<x<b_n$.

Now, let us consider the wave function in the first layer
which is given from \eqref{analypsi} by
\begin{eqnarray}\label{firstBarrier}
\psi_{1E}(x)&=&A {{2ik
e^{ikb_0}}\over{V_1(e^{p_1b_0}-A_1e^{-p_1b_0})}}\\
&&\left[
(-ik+p_1)e^{p_1x} +(ik+p_1)A_1 e^{-p_1x} \right],\nonumber
\end{eqnarray}
where $A_1$ is obtained through the step potential profile
of the resonant barrier structure.
The wave function given by Eq.~\eqref{firstBarrier} exponentially
increases with increasing $x$ when $|A_1|=0$; $A_1$ vanishes
at the energies of quasi-bound states. $|A_1|=0$ may be used as the
quantization condition for the energy levels of a particle confined
in a potential well. In the practical calculations,
transmission resonances occur at the energies where $|A_1|$ takes a
minimum value in a sudden dip, if $b_1-b_0$ and $b_n-b_{n-1}$ are
large enough. Such resonances indicate quasi-bound states.

Now, a potential well can be treated by uplifting the potential
function restricted on the region of the well and its surroundings
to form a resonant barrier structure that can be handled in our
approach. This resonant structure is used as a model potential of
the potential well under consideration. Then, each resonance
occurred in this model-potential problem corresponds to an energy
eigenvalue for the potential well, and the real part or the
imaginary part of the quasi-bound-state function obtained for a
resonant energy is the eigenfunction of the bound state belonging to
the corresponding eigenvalue for the potential well within a
normalization constant.

\section{V. Some Examples}
\begin{figure}
\includegraphics[width=0.45\textwidth]{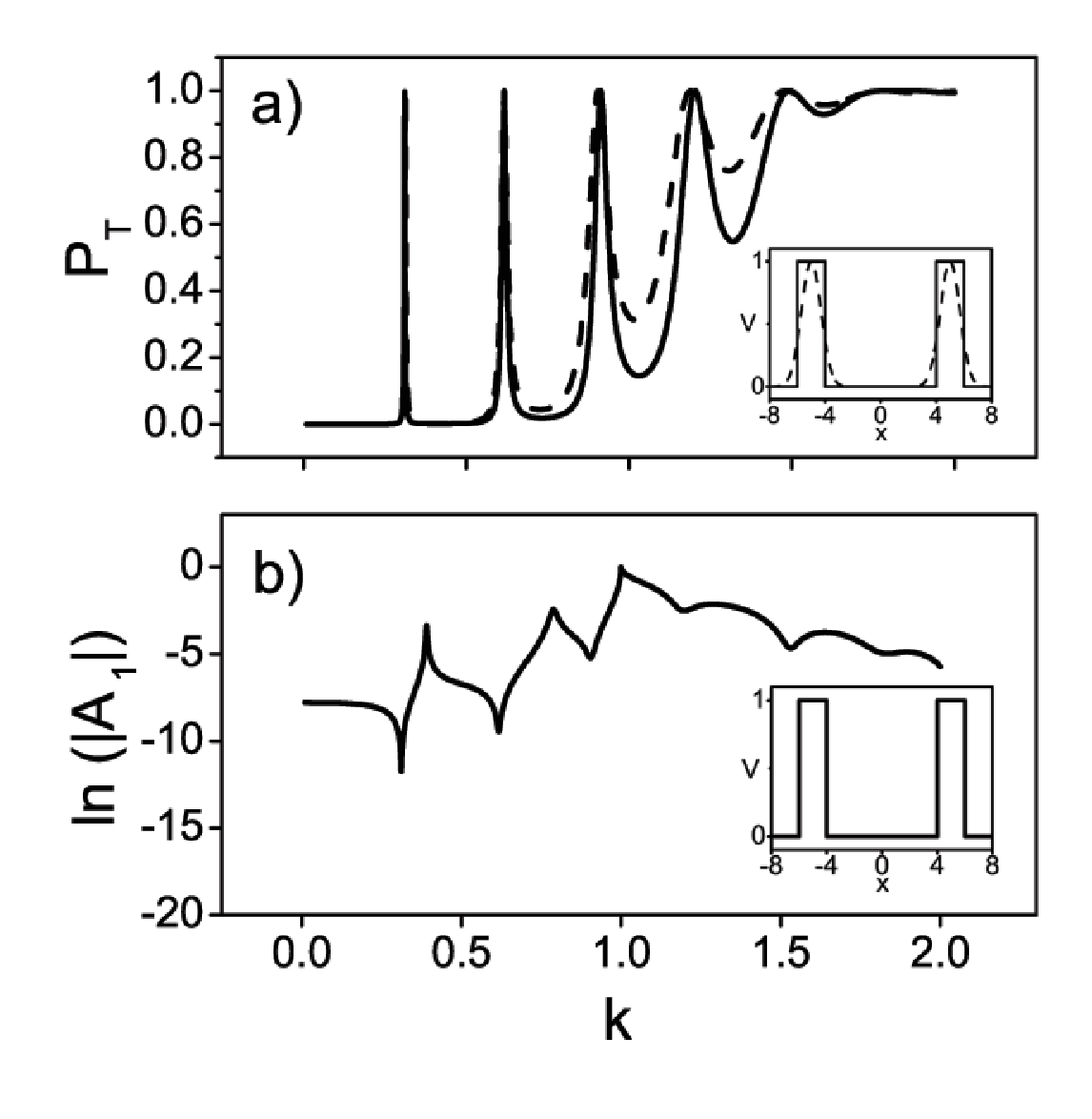}
\caption{\label{two}} (a) Transmission probabilities of
double-barrier potentials: the solid line and the dashed line are
for the square and the gaussian double-barrier potential,
respectively. Their potential profiles are shown in the inset graph.
(b) The curve of $\ln{(|A_1|)}$ for the double square-barrier
potential. Its profile is in the inset graph. Each dip of the $\ln
(|A_1|)$ curve below the barrier exactly agrees with the resonant
peak of  transmission probability shown in (a).
\end{figure}
In order to demonstrate the validity of our method, we consider
quantum barriers and quantum well potentials. Here, we examine the
derived recursive expressions of the transmission amplitude and the
state wave-function given by Eqs.~\eqref{Tpiecesolt} and
\eqref{analypsi}, respectively, by calculating $A_j$'s for several
potentials.

First, we consider rectangular and gaussian double-barrier
potentials which are shown in the inset of
Fig.~\protect\ref{two}(a). Here, the unit energy $\epsilon$ is
chosen to be the maximum potential energy of the barrier. Since the
rectangular double barrier structure consists of three layers of
constant potential, the analytic expression of the recursive
solution $\psi_{jE}(x)$ for this structure can be obtained with
calculated $A_j$'s which are dependent on $E$. One can determine
$A_3$ from Eq.~\eqref{AjCjp1} with the cutoff reflection amplitude
$C_4=R_E(x)=0$ in the region $x>b_3$, and determine $C_2$ for the
step region of $j=2$ where $V_2=0$ by evaluating $R_E(x)$ at $x=b_2$
in the step region of $j=3$ through Eq.~\eqref{Rpiecesol}. Then,
$A_1$ can be also obtained through Eq.~\eqref{AjCjp1} using the
value of $C_2$. Fig.~\protect\ref{two}(b) shows the curve of
$\ln(|A_1|)$ for the rectangular double barrier structure. Each dip
of the curve coincides with a transmission resonance for the energy
below the barrier ($k<1$) as shown in Fig.~\protect\ref{two}. Such a
sharp dip of $A_1$ indicates that an exponentially increasing term
of the wave function is dominant in the first barrier region due to
a negligibly small $|A_1|$ as was considered below
Eq.~\eqref{firstBarrier}, so that the transmission probability
becomes large in the narrow energy region. The resultant
transmission probabilities for the two barrier profiles have similar
resonance patterns as shown in Fig.~\protect\ref{two}(a). We have
confirmed that the solid line for the rectangular double barrier
perfectly agrees with the result obtained by the traditional method
for solving the Schr\"{o}dinger's equation for step potential
problems with boundary conditions imposed.

\begin{figure}
\includegraphics[width=0.45\textwidth]{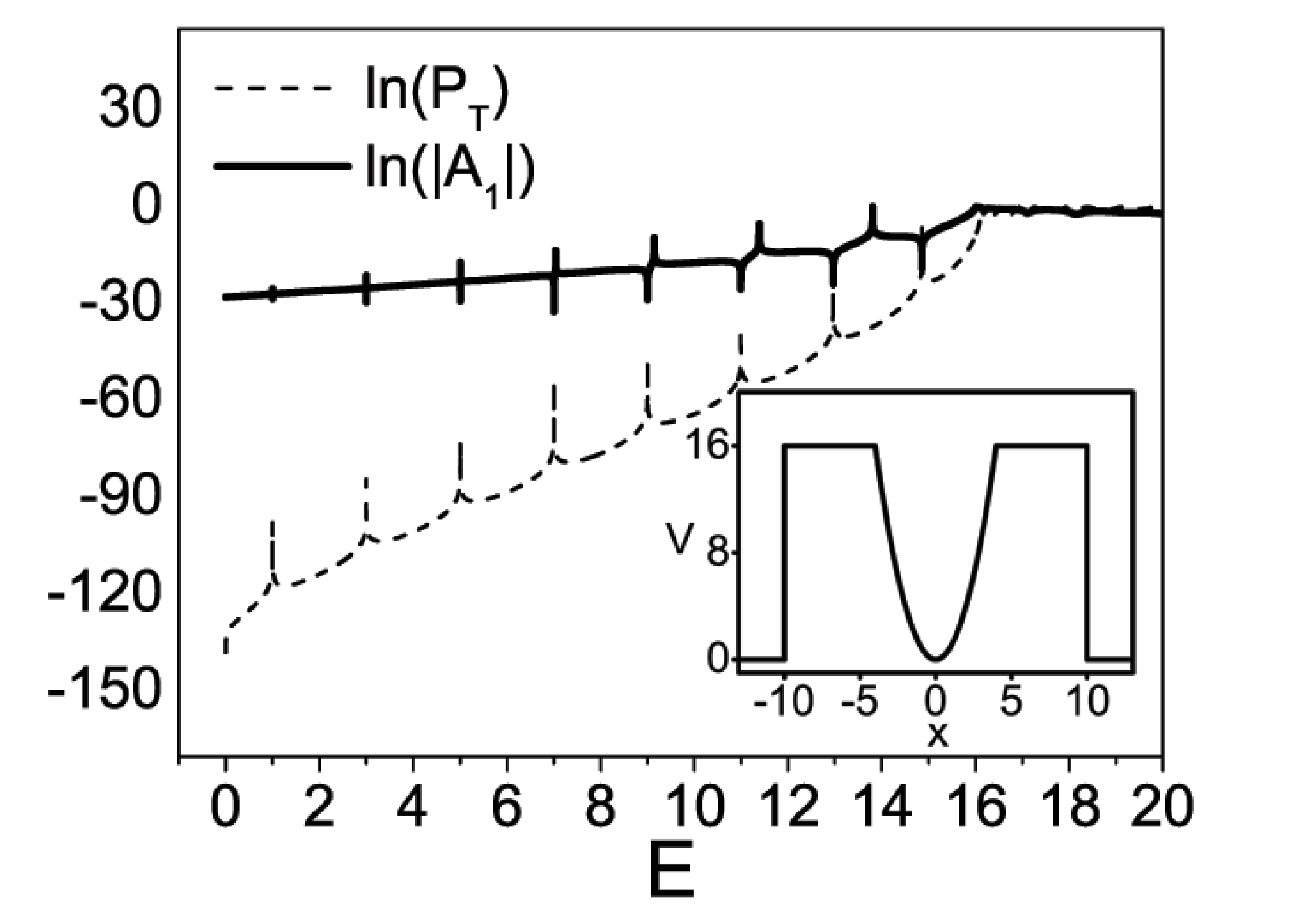}%
\caption{\label{three}} Resonances for a model potential of harmonic
oscillator given by Eq.~\protect\eqref{harmonic}. Each sharp dips on
the solid line of the curve $\ln{(|A_1|)}$ exactly agrees with a
resonant peak on the dashed line of the curve $\ln{(P_T)}$.
\end{figure}

Next, we consider a potential model by which one can obtain
eigenvalues and eigenfunctions for a harmonic oscillator. We take
the unit energy to be $\epsilon=\hbar\omega/2$, where $\omega$ is
the characteristic angular frequency of the harmonic oscillator.
Then, the dimensionless potential in Eq.~\eqref{Schr} is
$V_{ho}=x^2$ for the harmonic oscillator. In order to apply our
method to the harmonic oscillator in an energy range $0<E<16$, we
employ a model potential:
\begin{eqnarray}
\label{harmonic}
V_{HO}(x)=\left\{ \begin{array}{ll} x^2  &\mbox{for $|x|<4$}\\
  16  & \mbox{for $4<|x|<10$}\\
  0   & \mbox{for $|x|>10$},
  \end{array}
  \right.
\end{eqnarray}
as shown in the inset of Fig.~\protect\ref{three}. This model has a
double barrier structure and represents a quasi-bound problem in
which the transmission of a particle through the model potential
reaches its peak when the energy is resonant with one of the energy
levels of the quasi-bound states in the region between the two
potential barriers. The potential in Eq.~\eqref{harmonic} is
expected to involve low quasi-bound-state energies that
approximately equal the bound state energies of the harmonic
oscillator, respectively. Each sharp dip of $\ln(|A_1|)$ suppresses
the term exponentially decreasing with increasing $x$ in the wave
function in Eq.~\eqref{analypsi} allowing the exponentially
increasing term to be dominant in the first barrier, and thus occurs
at the same energy as a resonance peak which locates a quasi-bound
state, as shown in Fig.~\protect\ref{three}. Energies of the
quasi-bound states have been determined very accurately to fit in
the exact eigenvalues of the harmonic oscillator $E_n=2n+1$ with
$n=0,1,2,\cdots$ for the low lying states. However, there are very
small errors for the upper levels, for instance, $n=5,6,7$, because
the model potential in Eq.~\eqref{harmonic} is different from the
potential of the harmonic oscillator for the region $|x|>4$.

\begin{figure}
\includegraphics[width=0.45\textwidth]{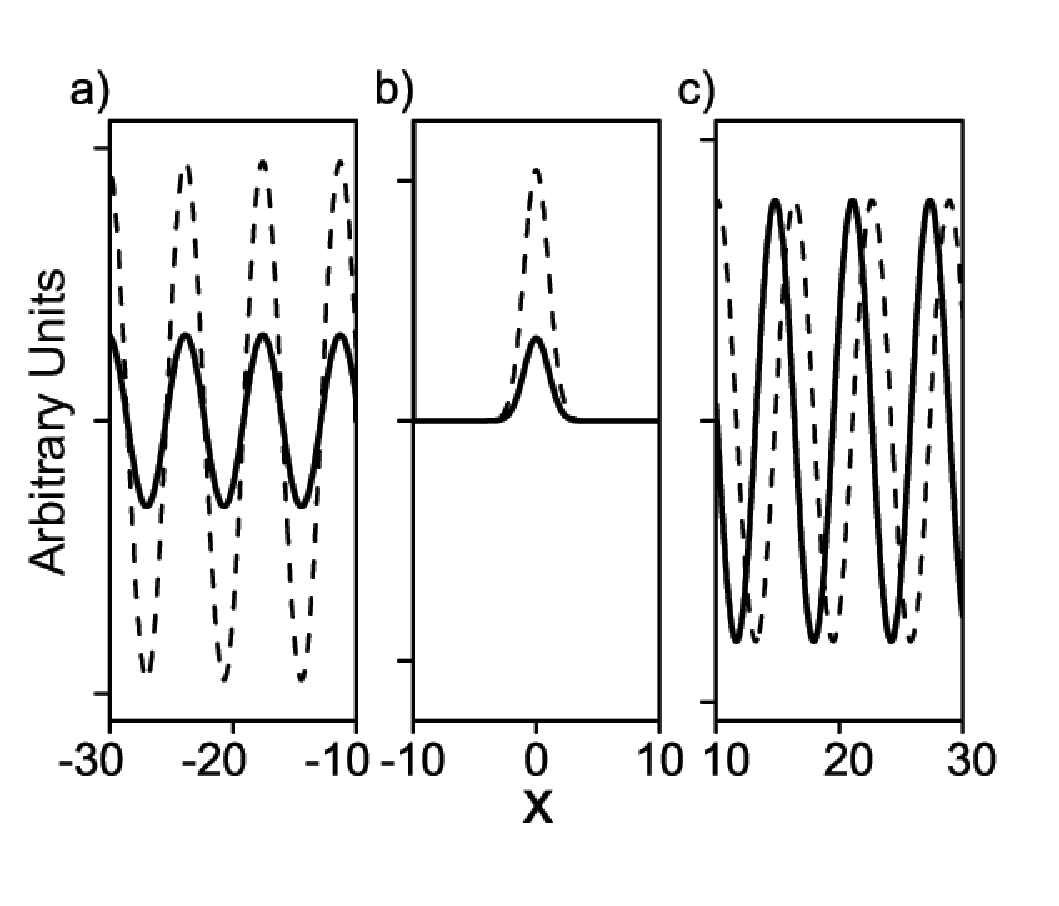}%
\caption{\label{four}} The wave function of the lowest quasi-bound
state ($E=1.0$) for the potential given by
Eq.~\protect\eqref{harmonic} for three regions with different units.
The solid line and dashed line are the real and imaginary parts of
the wave function, respectively.
\end{figure}

\begin{figure}
\includegraphics[width=0.45\textwidth]{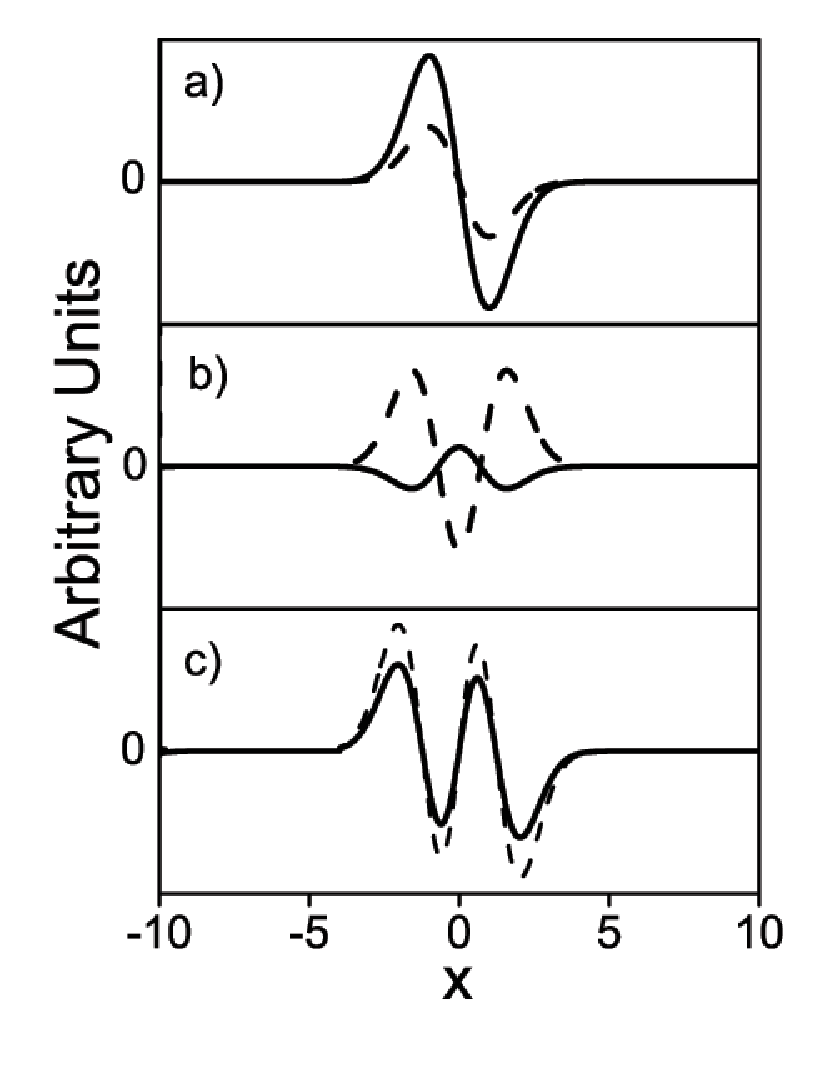}%
\caption{\label{five}} The wave functions of three quasi-bound
states of energies, (a) $E=3.0$, (b) $E=5.0$ and (c) $E=7.0$, for
the potential given by Eq.~\protect\eqref{harmonic}. The solid lines
and dashed lines are the real and imaginary parts of the wave
functions, respectively.
\end{figure}

Once a quasi-bound-state eigenvalue is determined, the wave function
of the state can be obtained from Eq.~\eqref{analypsi}. The lowest
quasi-bound state for the model potential in Eq.~\eqref{harmonic}
corresponds to the ground state for the harmonic oscillator. The
wave function of the state is separately shown  in
Fig.~\protect\ref{four} for the regions, $-30<x<-10$, $-10<x<10$,
and $10<x<30$ with different units. The solid line and dashed line
are, respectively, the real part and the imaginary part of the wave
function. The forward and reflected plane waves are superposed in
the region $x<-10$, and only the transmitted forward plane-wave
propagates in the region $x>10$. One can see in
Fig.~\protect\ref{four}(b) that the wave function exponentially
increases with increasing $x$ in the first barrier region at the
resonance energy but exponentially decreases in the second barrier
region which is sufficiently wide as explained earlier, forming the
lowest quasi-bound-state wave function in the binding potential
region between the two barriers. Wave functions of the next three
quasi-bound states for the same model potential that correspond to
the three lowest excited states in the harmonic oscillator problem
are also shown in Fig.~\protect\ref{five}. Either the real part or
the imaginary part of a quasi-bound-state wave function can be taken
as the wave function of the corresponding bound state for the
harmonic oscillator.

\begin{figure}
\includegraphics[width=0.45\textwidth]{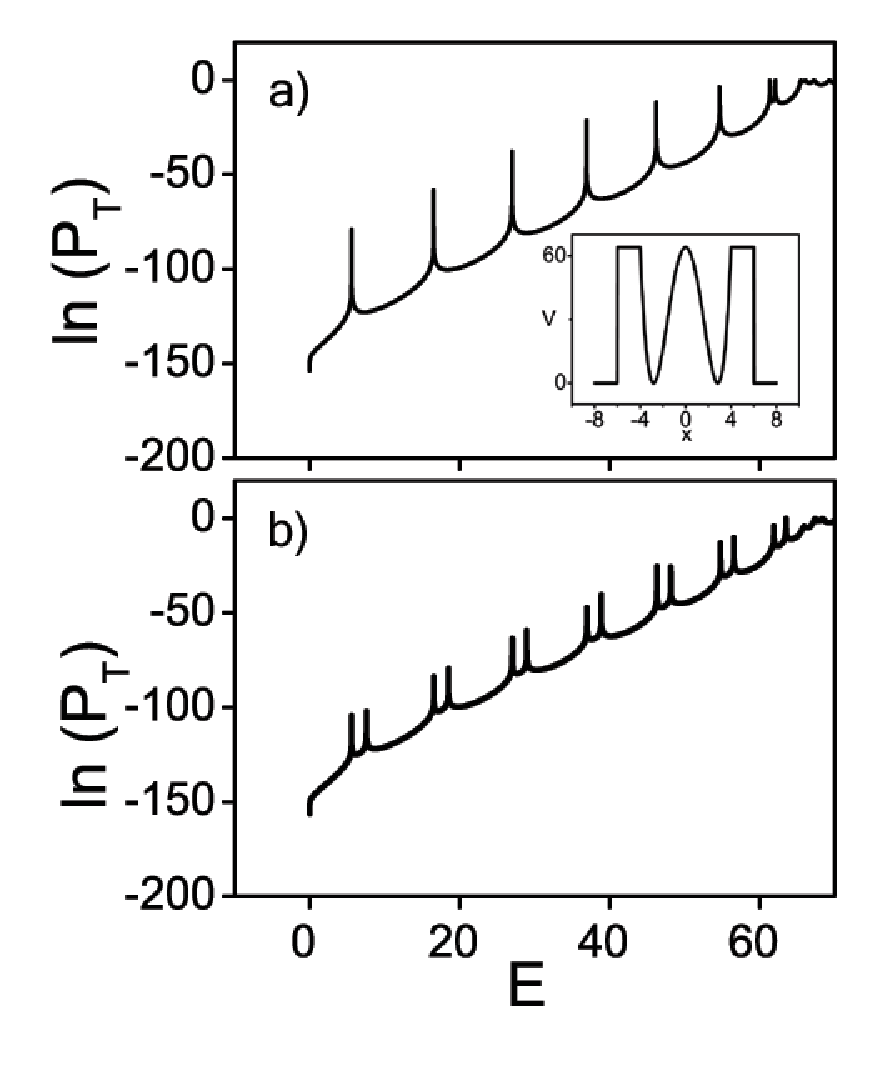}%
\caption{\label{six}} Logarithmic transmission probabilities for
double-well model potentials given by (a)
Eq.~\protect\eqref{doublewell} and (b)
Eq.~\protect\eqref{asymdouble}. Each peak represents transmission
resonance.
\end{figure}

For our final example, we consider a symmetric double-well potential
given by
\begin{eqnarray}
\label{doublewell0}
V_{dw}(x)=\left\{ \begin{array}{ll} x^2(x^2-16)  &\mbox{for $|x|<4$},\\
  0   & \mbox{for $|x|>4$}.
  \end{array}
  \right.
\end{eqnarray}
For treating this problem with our methods, we take a model
potential of $V_{dw}$ by uplifting the potential function on the
region $|x|<6$ by $64$:
\begin{eqnarray}
\label{doublewell}
V_{DW}(x)=\left\{ \begin{array}{ll} x^2(x^2-16)+64  &\mbox{for $|x|<4$},\\
  64  & \mbox{for $4<|x|<6$},\\
  0   & \mbox{for $|x|>6$},
  \end{array}
  \right.
\end{eqnarray}
which is shown in the inset of Fig.~\protect\ref{six}(a). The local
maximum of the model potential located at $x=0$ is $64$, and the two
local minima are located at $x=2\sqrt{2}$. The calculated curve for
$\ln{P_T}$ is plotted in Fig.~\protect\ref{six}(a). As is
well-known, each peak comprises a pair of states, an even state and
an odd state, whose energies are the very same especially for a peak
of lower energy. But they are not perfectly degenerate, since a
particle in one potential well can tunnel into the other well
through the middle barrier which is finite. The energy difference of
two states comprised in a peak increases with increasing energy, and
thus the seventh pair of states which are just below the barrier
appears to be separated.

\begin{figure}
\includegraphics[width=0.45\textwidth]{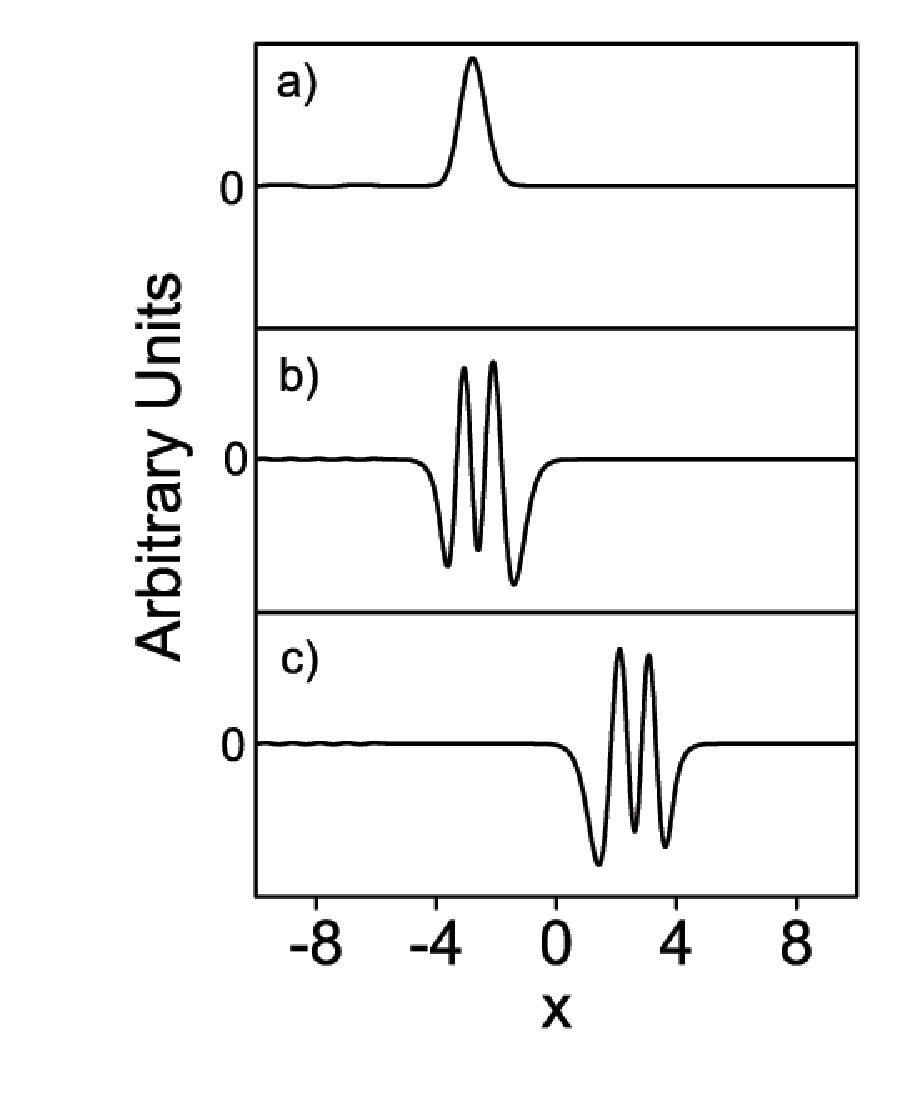}%
\caption{\label{seven}} The wave functions of three bound states for
the double well potential, $V(x)=x^2(x^2-16)+\tanh x$;
 (a) the ground state, (b) the eighth excited state, and
(c) the ninth excited state. These wave functions have been obtained
by using the model potential given by Eq.~\protect\eqref{asymdouble}
as explained in the text.
\end{figure}

In order to confirm that two states are comprised in a peak, we add
a small asymmetric potential function to the potential in
Eq.~\eqref{doublewell}:
\begin{equation}
\label{asymdouble} V'_{DW}(x)=V_{DW}(x)+ V'(x)=V_{DW}(x) + \tanh
x+1.
\end{equation}
Using this potential, we obtain a curve of $\ln |P_T|$ in which
there are seven pairs of quasi-bound states as shown in Fig.~6(b).
Each pair is split into two peaks due to the perturbing potential
$V'(x)$ from a peak containing two states for the double-well
potential $V_{DW}$ shown in Fig.~6(a). The lower state and higher
state in each pair are expected to be confined in the left well and
right well, respectively.

In order to find bound-state wave functions for the double well
potential given as $V(x)=x^2(x^2-16)+\tanh x$, one may take either
the real parts or imaginary parts of quasi-bound-state wave
functions determined from Eq.~\eqref{analypsi} for an appropriate
model potential such as Eq.~\eqref{asymdouble}. The real parts (or
imaginary parts) can be considered as the bound-state
eigenfunctions within a normalization constant. Real parts of the
wave function of the lowest ($n=0$), the ninth ($n=8$), and the
tenth ($n=9$) quasi-bound states, which are taken to be the wave
functions of corresponding bound states, are plotted in Fig.~7(a),
(b) and (c), respectively. Fig.~7 shows that the particle which
occupies the lowest state or the ninth state stays in the left
potential well, while the particle occupying the tenth state stays
in the right potential well, as expected. 

\section{VI. Discussions and Conclusion}
Analytically solving two nonlinear first-order differential
equations equivalent to the Schr\"{o}dinger's equation, we have
obtained the recursive solutions of the cutoff reflection amplitude
$R_E(x)$, the cutoff transmission amplitude $T_E(x)$ and the state
wave function $\psi_E(x)$  for a multi-step potential. One can use
the recursive solutions in Eqs.~\eqref{Tpiecesolt} and
\eqref{analypsi} for any piecewise constant potential to obtain
transmission amplitudes and eigenfunctions, respectively. If an
approximate multi-step potential consisting of sufficiently many
layers is substituted for a smooth potential profile, calculations
with the recursive solutions are accurate and fast.

For treating a potential well, we take a model potential with two
side barriers that contains a partial profile identical to the
profile of the potential well in the region between the side
barriers. In the model potential problem, resonance occurs at the
energy of a quasi-bound state, so the energies can be determined by
locating the transmission resonances. Once a quasi-bound-state
energy is known, the eigenfunction for the energy can also be
determined. However, the real part and the imaginary part of the
wave function of a quasi-bound state exponentially increases at a
same rate in the first barrier and exponentially decrease in the
last barrier as explained earlier, and are the same within a factor
$\tan\phi$ in the region between two side barriers, $\phi$ being a
phase angle. Therefore, either the real or the imaginary part can be
taken for the wave function of the corresponding bound state with a
normalization constant.

We have considered several examples to demonstrate the validity of
our method with the recursive solutions. First, we have calculated
the transmission probabilities of a particle incident upon a
rectangular and a gaussian double-barrier potential which show
typical features of resonant tunneling in Fig.~\protect\ref{two}(a).
The rectangular double barrier simply consists of three layers, the
left barrier being the first potential step, and the feature of
resonant tunneling can be explained by the calculational results of
$A_1$ in which information of the barrier structure is included. If
$A_1=0$ for an energy lower than the barriers, the wave function
does not contain exponentially decreasing term with increasing $x$
in the first barrier region so that it exponentially increases in
the region, and thus the transmission probability reaches maximum at
the energy. Practically, the sharp dips of the $\ln |A_1|$ curve in
Fig.~2(b) represent the resonances associated with quasi-bound
states in the rectangular well between the barriers.

The harmonic oscillator problem can be dealt with by solving a
quasi-bound-state problem with the model potential given by
Eq.~\eqref{harmonic}. The calculational results for $\ln |A_1|$ and
$\ln P_T$ are shown in Fig.~\protect\ref{three}, so the
quasi-bound-state energies which may be regarded as the energy
eigenvalues of the harmonic oscillator can readily be determined by
locating the resonance peaks of $\ln P_T$ or the sharp dips of $\ln
|A_1|$. The resultant peaks perfectly agree with the exact
eigenvalues for low energy states, but deviate a little from the
exact values for high energy states; $E_0=1.000$, $E_1=3.000$,
$E_2=5.000$, $E_3=6.9999$ $E_4=8.999$, $E_5=10.994$, $E_6=12.970$,
and $E_7=14.857$. Since the deviation originates from the
approximated model potential which does not agree with the
harmonic-oscillator potential on the high potential region $|x|>4$,
one may get more accurate eigenvalues for higher states by using a
larger model potential which coincides with the harmonic-oscillator
potential in a wider region.

We have applied our method to a symmetric double-well potential
given by Eq.~\eqref{doublewell0}, using Eq.~\eqref{doublewell} as
its model potential. It is well-known that an even state and an odd
state are folded together belonging closely to an eigenvalue if they
are deeply bound in a symmetric double well whose middle barrier is
strong enough so that the wave functions nearly vanish in the barrier
region. The even state and odd state belonging to the same eigenvalue can
be combined to constitute two bound states that are localized in the
left well and right well with the same eigenvalue, respectively.
Thus, one may consider each resonance peak to correspond to a pair of states
confined in one and the other well, as well as a pair of even and
odd states. When an asymmetric perturbing potential is added to 
the double-well potential, each resonance peak has been split into two peaks, 
as shown in Fig.~\protect\ref{six}(b). Since the potential
is not symmetric due to the perturbation, even and odd states can
not be sustained. Therefore, each resonance peak in
Fig.~\protect\ref{six}(b) represents a state confined in one of the
two wells. As expected, Fig.~\protect\ref{seven} shows that among
the two states separated from each other due to the perturbation, the
lower energy state is confined in the left well which is a little deeper
than the right well, while the higher energy state is confined in the
right-well region. The ninth (n=8) and tenth (n=9) states are the fifth states in the left well and the right well, respectively, so the numbers of nodes in their wave functions are identically four as in Fig 7 (b) and (c). 

Here, more accurate determination of
quasi-bound-state energies is required to obtain eigenfunctions of
the lower quasi-bound-state confined in the right well, because the
resonance is sharper for the particle tunneling into a deeper
quasi-bound state or a state confined in the right well through the
stronger barrier or two barriers (the first barrier and the middle
barrier); for instance, the determined eigenvalues
$E_0=5.601849104$, $E_1=7.58342367856952$, $E_8=46.290706$ and
$E_9=48.1538536$ should be used in obtaining the wave functions,
where $E_1$ is required to be the most accurate. Here, each
bound-state energy for $V(x)=x^2(x^2-16)+\tanh x$ is the
corresponding quasi-bound-state energy for $V'_{DW}$ minus 65.
Unlike the model potential of the harmonic oscillator, no
approximation is involved in the calculation with the model
potential of this double well potential.

Although here we have presented results for double barriers, a
harmonic oscillator, and double-well potentials, the analysis can
easily be applied for accurate calculations to general potential
barrier and quantum well structures. The analysis could be extended
to the one-dimensional problem of a particle that has a
coordinate-dependent mass, by imposing the boundary conditions on
the step potential structure. In addition, the three-dimensional
Schr\"{o}dinger equation for a spherically symmetric potential can
also be solved using the present method because such a
three-dimensional problem can be reduced to a one-dimensional
problem.

In conclusion, we have described a method for solving the
one-dimensional Schr\"{o}dinger equation by means of the analytic
expressions of recursive solutions derived for piecewise-constant
potentials. The recursive solutions provide a general analysis of
one-dimensional scattering, quasi-bound and bound state problems. We
have demonstrated the validity of the method by taking some examples
to show consistent and predictable calculational results. The
calculations with the recursive solutions were rapid and accurate
even for smoothly varying potentials.

\subsection{}

\begin{acknowledgments}
This work was financially supported by Dankook University (2005
research fund).
\end{acknowledgments}

\bibliography{ref}

\begin{thebibliography}{14}
\expandafter\ifx\csname natexlab\endcsname\relax\def\natexlab#1{#1}\fi
\expandafter\ifx\csname bibnamefont\endcsname\relax
  \def\bibnamefont#1{#1}\fi
\expandafter\ifx\csname bibfnamefont\endcsname\relax
  \def\bibfnamefont#1{#1}\fi
\expandafter\ifx\csname citenamefont\endcsname\relax
  \def\citenamefont#1{#1}\fi
\expandafter\ifx\csname url\endcsname\relax
  \def\url#1{\texttt{#1}}\fi
\expandafter\ifx\csname urlprefix\endcsname\relax\def\urlprefix{URL }\fi
\providecommand{\bibinfo}[2]{#2}
\providecommand{\eprint}[2][]{\url{#2}}

\bibitem[{\citenamefont{Barnham and Vvedensky}(2001)}]{Barnham}
\bibinfo{editor}{\bibfnamefont{K.}~\bibnamefont{Barnham}} \bibnamefont{and}
  \bibinfo{editor}{\bibfnamefont{D.}~\bibnamefont{Vvedensky}}, eds.,
  \emph{\bibinfo{title}{Low-dimensional Semiconductor Structure,}}
  (\bibinfo{publisher}{Cambridge University Press},
  \bibinfo{address}{Cambridge}, \bibinfo{year}{2001}).

\bibitem[{\citenamefont{Brennan and Summers}(1987)}]{Brennan}
\bibinfo{author}{\bibfnamefont{K.~E.} \bibnamefont{Brennan}} \bibnamefont{and}
  \bibinfo{author}{\bibfnamefont{C.~J.} \bibnamefont{Summers}},
  \bibinfo{journal}{J. Appl. Phys.} \textbf{\bibinfo{volume}{61}},
  \bibinfo{pages}{614} (\bibinfo{year}{1987}).

\bibitem[{\citenamefont{Ghatak et~al.}(1988)\citenamefont{Ghatak, Thyagarajan,
  and Shenoy}}]{Ghatak}
\bibinfo{author}{\bibfnamefont{A.~K.} \bibnamefont{Ghatak}},
  \bibinfo{author}{\bibfnamefont{K.}~\bibnamefont{Thyagarajan}},
  \bibnamefont{and} \bibinfo{author}{\bibfnamefont{M.~R.}
  \bibnamefont{Shenoy}}, \bibinfo{journal}{IEEE J. Quantum Electron.}
  \textbf{\bibinfo{volume}{24}}, \bibinfo{pages}{1524} (\bibinfo{year}{1988}).

\bibitem[{\citenamefont{Jonsson and Eng}(1990)}]{Jonsson}
\bibinfo{author}{\bibfnamefont{B.}~\bibnamefont{Jonsson}} \bibnamefont{and}
  \bibinfo{author}{\bibfnamefont{S.~T.} \bibnamefont{Eng}},
  \bibinfo{journal}{IEEE J. Quantum Electron.} \textbf{\bibinfo{volume}{26}},
  \bibinfo{pages}{2025} (\bibinfo{year}{1990}).

\bibitem[{\citenamefont{Anemogiannis et~al.}(1999)\citenamefont{Anemogiannis,
  Glytsis, and Gaylord}}]{Glytsis}
\bibinfo{author}{\bibfnamefont{E.}~\bibnamefont{Anemogiannis}},
  \bibinfo{author}{\bibfnamefont{E.~N.} \bibnamefont{Glytsis}},
  \bibnamefont{and} \bibinfo{author}{\bibfnamefont{T.~K.}
  \bibnamefont{Gaylord}}, \bibinfo{journal}{Microelectron. J.}
  \textbf{\bibinfo{volume}{30}}, \bibinfo{pages}{935} (\bibinfo{year}{1999}).

\bibitem[{\citenamefont{Rakityansky}(2004)}]{Rakityansky}
\bibinfo{author}{\bibfnamefont{S.~A.} \bibnamefont{Rakityansky}},
  \bibinfo{journal}{Phys. Rev. B} \textbf{\bibinfo{volume}{70}},
  \bibinfo{pages}{205323} (\bibinfo{year}{2004}).

\bibitem[{\citenamefont{Singh}(1986)}]{Singh}
\bibinfo{author}{\bibfnamefont{J.}~\bibnamefont{Singh}},
  \bibinfo{journal}{Appl. Phys. Lett.} \textbf{\bibinfo{volume}{48}},
  \bibinfo{pages}{434} (\bibinfo{year}{1986}).

\bibitem[{\citenamefont{Nakamura et~al.}(1989)\citenamefont{Nakamura, Shimizu,
  Koshiba, and Hayata}}]{Nakamura}
\bibinfo{author}{\bibfnamefont{K.}~\bibnamefont{Nakamura}},
  \bibinfo{author}{\bibfnamefont{A.}~\bibnamefont{Shimizu}},
  \bibinfo{author}{\bibfnamefont{M.}~\bibnamefont{Koshiba}}, \bibnamefont{and}
  \bibinfo{author}{\bibfnamefont{K.}~\bibnamefont{Hayata}},
  \bibinfo{journal}{IEEE J. Quantum Electron.} \textbf{\bibinfo{volume}{25}},
  \bibinfo{pages}{889} (\bibinfo{year}{1989}).

\bibitem[{\citenamefont{Love and Winkler}(1977)}]{Love}
\bibinfo{author}{\bibfnamefont{J.~D.} \bibnamefont{Love}} \bibnamefont{and}
  \bibinfo{author}{\bibfnamefont{C.}~\bibnamefont{Winkler}},
  \bibinfo{journal}{J. Opt. Soc. Am.} \textbf{\bibinfo{volume}{67}},
  \bibinfo{pages}{1627} (\bibinfo{year}{1977}).

\bibitem[{\citenamefont{Ghatak et~al.}(1992)\citenamefont{Ghatak, Gallawa, and
  Gayal}}]{Ghatak0}
\bibinfo{author}{\bibfnamefont{A.~K.} \bibnamefont{Ghatak}},
  \bibinfo{author}{\bibfnamefont{R.~L.} \bibnamefont{Gallawa}},
  \bibnamefont{and} \bibinfo{author}{\bibfnamefont{I.~C.} \bibnamefont{Gayal}},
  \bibinfo{journal}{IEEE J. Quantum Electron.} \textbf{\bibinfo{volume}{28}},
  \bibinfo{pages}{400} (\bibinfo{year}{1992}).

\bibitem[{\citenamefont{Tikochinsky}(1977)}]{Tikochinsky}
\bibinfo{author}{\bibfnamefont{Y.}~\bibnamefont{Tikochinsky}},
  \bibinfo{journal}{Ann. Phys.} \textbf{\bibinfo{volume}{103}},
  \bibinfo{pages}{185} (\bibinfo{year}{1977}).

\bibitem[{\citenamefont{Goodvin and Shegelski}(2005{\natexlab{a}})}]{Glen1}
\bibinfo{author}{\bibfnamefont{G.~L.} \bibnamefont{Goodvin}} \bibnamefont{and}
  \bibinfo{author}{\bibfnamefont{M.~R.~A.} \bibnamefont{Shegelski}},
  \bibinfo{journal}{Phys. Rev. A} \textbf{\bibinfo{volume}{71}},
  \bibinfo{pages}{032719} (\bibinfo{year}{2005}{\natexlab{a}}).

\bibitem[{\citenamefont{Goodvin and Shegelski}(2005{\natexlab{b}})}]{Glen2}
\bibinfo{author}{\bibfnamefont{G.~L.} \bibnamefont{Goodvin}} \bibnamefont{and}
  \bibinfo{author}{\bibfnamefont{M.~R.~A.} \bibnamefont{Shegelski}},
  \bibinfo{journal}{Phys. Rev. A} \textbf{\bibinfo{volume}{72}},
  \bibinfo{pages}{042713} (\bibinfo{year}{2005}{\natexlab{b}}).

\bibitem[{\citenamefont{Razavy}(2003)}]{Razavy}
\bibinfo{author}{\bibfnamefont{M.}~\bibnamefont{Razavy}},
  \emph{\bibinfo{title}{Qunatum Theory of Tunneling}}
  (\bibinfo{publisher}{World Scientific}, \bibinfo{address}{Singapore},
  \bibinfo{year}{2003}).

\end{thebibliography}

\end{document}